\documentclass[conference]{IEEEtran}
\IEEEoverridecommandlockouts
\usepackage{amsthm}
\usepackage{amsmath}
\usepackage{latexsym}
\usepackage{graphicx}
\usepackage{bbding}
\usepackage{indentfirst}
\usepackage{algorithm,algorithmic}
\usepackage{setspace}
\usepackage{float}
\usepackage{epstopdf}
\usepackage{amssymb}
\usepackage{amsfonts}
\usepackage{enumerate}
\usepackage{multicol}
\usepackage{color}

\newcommand{\mbf}[1]{\mathbf{#1}}
\newcommand{\mbb}[1]{\mathbb{#1}}
\newcommand{\mc}[1]{\mathcal{#1}}
\newcommand{\expa}[1]{\mathbb{E}\left[ #1 \right]}

\newcommand{\trace}[1]{\text{tr}\left(#1\right)}

\begin{document}
	
	\title{Low-Complexity 6DMA Rotation and Position Optimization Based on Statistical Channel Information}
	
		\author{\IEEEauthorblockN{Qijun Jiang$\IEEEauthorrefmark{1}$,
				Xiaodan Shao$\IEEEauthorrefmark{2}$, and Rui Zhang$\IEEEauthorrefmark{1}$$\IEEEauthorrefmark{3}$}
		\IEEEauthorblockA{$\IEEEauthorrefmark{1}$School of Science and Engineering, The Chinese University of Hong Kong, Shenzhen, China\\
			$\IEEEauthorrefmark{2}$Department of Electrical and Computer Engineering, University of Waterloo, Canada\\
			$\IEEEauthorrefmark{3}$Department of Electrical and Computer Engineering, National University of Singapore, Singapore\\
			E-mails: qijunjiang@link.cuhk.edu.cn, x6shao@uwaterloo.ca, elezhang@nus.edu.sg}\vspace{-23pt}	
			
			\thanks{This work is supported in part by The Guangdong Provincial Key Laboratory of Big Data Computing, the National Natural Science Foundation of China (No. 62331022), and the Guangdong Major Project of  Basic and Applied Basic Research (No.  2023B0303000001).}
	}\maketitle

	\IEEEpeerreviewmaketitle
\begin{abstract}
The six-dimensional movable antenna (6DMA) is a promising technology to fully exploit spatial variation in wireless channels by allowing flexible adjustment of three-dimensional (3D) positions and rotations of antennas at the transceiver. In this paper, we consider a 6DMA-equipped base station (BS) and aim to  maximize the average sum logarithmic rate of all users served by the BS  by jointly designing 6DMA surface positions and rotations based on statistical channel  information (SCI). Different from prior works on  6DMA design which  use alternating optimization to iteratively update surface positions and rotations, we propose a new  sequential optimization method that first determines the optimal rotations and then identifies feasible positions to realize these rotations under practical antenna placement constraints. Simulation results show that our proposed optimization scheme  significantly reduces the computational complexity of conventional alternating optimization (AO), while achieving communication performance comparable to the AO-based approach and superior to  existing fixed-position/rotation antenna arrays.
 \end{abstract}


\section{Introduction}
To accommodate the escalating demand for higher data rates driven by a growing population of wireless devices in upcoming sixth-generation (6G) networks \cite{10494370},  multiple-input and multiple-output (MIMO) technologies are trending toward equipping transceivers with a dramatically larger number of antennas \cite{10045774,ruiIRS,10858129,10555049}. However, boosting performance by merely increasing fixed-position antennas entails greater hardware costs and energy consumption. Moreover, systems with fixed-position antennas cannot flexibly allocate antenna resources according to users’ spatial distributions, which leads to inefficient antenna utilization.

To solve these challenges, six-dimensional movable antenna (6DMA) has been proposed as a promising technology to fully exploit spatial variation in wireless channels by allowing flexible adjustment of three-dimensional (3D) positions and rotations of antennas at the transceiver \cite{6DMA_TWC,6DMA_JSAC}. 6DMA can adapt to the spatial channel distribution of users in a given environment and realize enhanced communication performance with intelligent reconfiguration. Recent studies have demonstrated the potential of 6DMA in various applications \cite{s2025tutorial}. For example, the works in \cite{6DMA_TWC} and \cite{6DMA_JSAC} introduce the 6DMA architecture and formulate an ergodic capacity maximization problem, revealing significant performance gains from optimizing continuous and discrete antenna positions and rotations. Moreover, the authors in \cite{6DMA_JSTSP} propose low-training-complexity channel estimation algorithms by exploiting a new directional-sparsity characteristic of 6DMA channels. Subsequent works explore polarized 6DMA \cite{IPA}, cell-free 6DMA \cite{free6DMA, wen}, 6DMA-aided unmanned aerial vehicle (UAV) \cite{UAV6DMA,wang2025uav}, 6DMA-aided intelligent reflecting surface (IRS) \cite{passive6DMA}, hybrid near-far field 6DMA \cite{near}, and 6DMA-enhanced wireless sensing \cite{6dmasensing}. Compared with movable antennas with fixed rotations \cite{MA,ding2024movable}, 6DMA achieves superior performance with practical mechanical adaptation and low-frequency reconfiguration.

Despite the advances in 6DMA, its practical deployment still faces several challenges. A key difficulty lies in the non-convex optimization of 6DMA positions and rotations in a high-dimensional space. Existing approaches, such as alternating optimization (AO), rely on Monte Carlo sampling to approximate ergodic performance based on a priori known channel state information (CSI) \cite{6DMA_TWC,6DMA_JSAC}. These methods incur a high computational cost and may converge to undesired local optima. Moreover, because 6DMA surfaces can adapt only on a slow timescale due to their mechanical control mechanisms \cite{MA}, instantaneous CSI-based optimization is neither practical nor necessary. Fortunately, 6DMA can be optimized using users’ statistical channel information (SCI), which varies more slowly with user positions and scattering environments. This makes SCI-based optimization a more viable and energy-efficient solution for the practical deployment of 6DMA.

To address these challenges, we propose a low-complexity, sequential optimization approach for 6DMA based on users’ SCI. Instead of alternating between position and rotation updates, our method first determines the optimal rotation of each 6DMA surface and then finds feasible positions to realize these rotations under physical placement constraints. To avoid the complexity of Monte Carlo–based optimization, we derive an analytical approximation of the average achievable rate, which facilitates efficient evaluation of the 6DMA configuration. Simulation results demonstrate that the proposed sequential optimization achieves performance comparable to the benchmark AO-based approach, while significantly reducing computational cost. Moreover, it consistently outperforms conventional fixed-position antenna arrays and offers a practical solution for 6DMA system implementation in wireless networks.
\section{System Model and Problem Formulation}
\subsection{Channel Model}
\begin{figure}[!t] 
	\centering
	\setlength{\abovecaptionskip}{-3pt}
	\setlength{\belowcaptionskip}{-15pt}
	\includegraphics[width=3.1in]{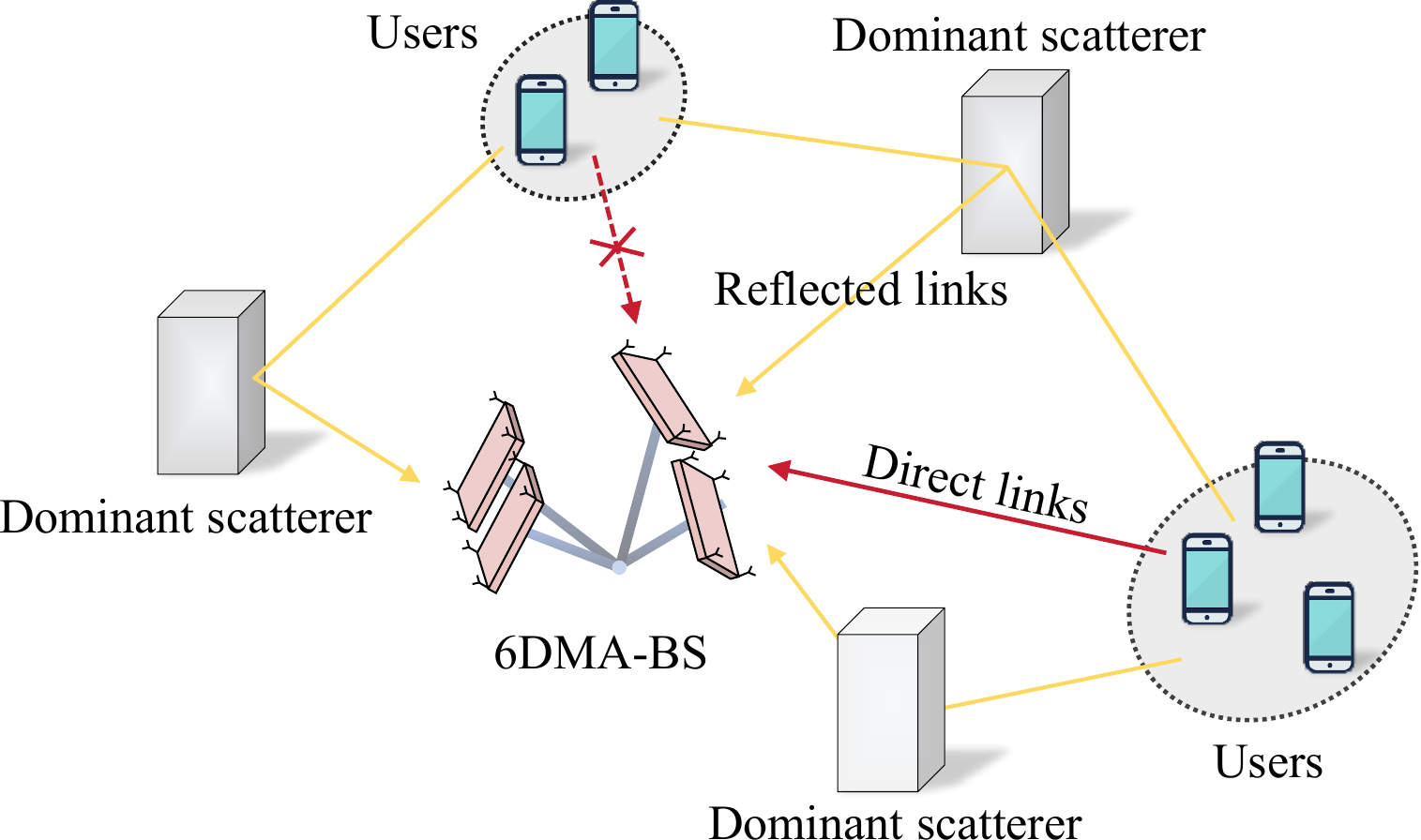} 
	\caption{6DMA-enabled wireless communication system.} 
	\label{fig:Channel_model} 
			\vspace{-0.58cm}
\end{figure}

We consider a 6DMA-aided communication system in which a single base station (BS) equipped with 6DMA serves $K$ spatially distributed users. As shown in Fig.\,\ref{fig:Channel_model},
each user $k$, $k \in \mc{K} \triangleq \{1,\cdots,K\}$, has  $L_{k}$ propagation  paths to the BS, indexed by the set $\mc{L}_k \triangleq \{1,\cdots,L_k\}$, which result in the direct link and/or  reflected links  by dominant scatterers. The 6DMA-BS is equipped with $B$ 6DMA surfaces, indexed by the set $\mc{B} \triangleq \{1,\cdots,B\}$,  each  comprising $N$  directional antennas, indexed by the set $\mc{N}\triangleq\{1,\cdots,N\}$. All users are each equipped with an omni-directional antenna. The positions and rotations (orientations) of all 6DMA surfaces can be individually adjusted. In particular, the position and rotation  of the $b$-th 6DMA surface, $b\in \mc{B}$, can be respectively characterized by
\begin{align}
	\mbf{q}_b \triangleq [x_b,y_b,z_b]^\top \in \mc{V}_{\text{6DMA}},~
	\mbf{u}_b \triangleq [\alpha_b, \beta_b, \gamma_b]^\top,
\end{align}
where $\mathcal{V}_{\text{6DMA}}$ denotes the given 3D region at the BS in which the 6DMA surfaces can be flexibly positioned/rotated. In the above, $x_b$, $y_b$, and $z_b$ represent the coordinates of the center of the $b$-th 6DMA surface in the global Cartesian coordinate system (CCS) $o$-$xyz$, with the 6DMA-BS’s reference position serving as the origin $o$; $\alpha_b $, $\beta_b $, and $\gamma_b $  denote the Euler angles with respect to (w.r.t.) the $x$-axis, $y$-axis and $z$-axis in the global CCS, respectively, in the range of $\left[0,2\pi\right)$.  
The end-to-end channels between the $K$ users and the 6DMA-BS  can be expressed as
\begin{align}
	\mbf{H}(\mbf{z}) \triangleq \begin{bmatrix} \mbf{h}_1(\mbf{z}_1) & \cdots & \mbf{h}_K(\mbf{z}_1)\\
		\vdots        &        & \vdots \\
		\mbf{h}_1(\mbf{z}_B) & \cdots & \mbf{h}_K(\mbf{z}_B)\\
	\end{bmatrix}  \in \mbb{C}^{BN\times K},
\end{align}
where $\mbf{z}_b \triangleq  [\mbf{q}_b^\top, \mbf{u}_b^\top]^\top \in \mbb{R}^{6}$ denotes the position-rotation pair of the $b$-th 6DMA surface, $\mbf{z} \triangleq [\mbf{z}_1^\top,\cdots,\mbf{z}_B^\top]^\top \in \mbb{R}^{6 B}$  denotes the position-rotation state of the 6DMA-BS, and $\mbf{h}_k(\mbf{z}_b) \in \mathbb{C}^{N}$ denotes the channel from the $k$-th user to the $b$-th 6DMA surface.  Consequently, $\mbf{h}_k(\mbf{z}_b) $ can be expressed as
\begin{align}
	\mbf{h}_k(\mbf{z}_b) &= \sum_{l=1}^{L_k} \sqrt{g(\mbf{u}_b,\mbf{f}_{k,l})}\mathbf{{a}}(\mbf{z}_b,\mbf{f}_{k,l})  v_{k,l}, \label{hkb}
\end{align}
where 
$\mbf{f}_{k,l}\in \mbb{R}^{3}$ and $v_{k,l}$, $k \in \mc{K}$, $l \in \mc{L}_k$, denote the direction of arrival (DoA) and complex channel coefficient of the $l$-th path of the $k$-th user, respectively;  $g(\mbf{u}_b,\mbf{f}_{k,l})$ 
and $\mathbf{{a}}(\mbf{z}_b,\mbf{f}_{k,l}) \in \mbb{C}^{N}$, $b \in \mc{B}$, respectively denote the  antenna gain and  steering vector of the $b$-th 6DMA surface. Specifically, we have
\begin{align}
	g(\mbf{u}_b,\mbf{f}_{k,l}) &= G\left(\mbf{R}^{-1}(\mbf{u}_b) \mbf{f}_{k,l}\right), \label{antenna_gain}\\
	\mathbf{{a}}(\mbf{z}_b,\mbf{f}_{k,l}) &= [e^{-j\frac{2\pi}{\lambda}\mbf{f}_{k,l}^{\top} \mbf{r}_{1}(\mbf{z}_b)},\cdots,e^{-j\frac{2\pi}{\lambda}\mbf{f}_{k,l}^{\top} \mbf{r}_{N}(\mbf{z}_b)}]^\top \label{steering_vec_def},
\end{align}
where $\mbf{R}(\mbf{u}_b)$  is the rotation matrix corresponding to rotation $\mbf{u}_b$, the function $G(\cdot)$ denotes the effective gain of antennas of each 6DMA surface in terms of the DoA in its local CCS, $\lambda$ denotes the wavelength of the carrier wave, and $\mbf{r}_n(\mbf{z}_b) \in \mbb{R}^3$ represents the location of the $n$-th antenna on the $b$-th 6DMA surface in the global CCS\cite{6DMA_TWC}.

Let $\mbf{h}_k(\mbf{z}) \triangleq [\mbf{h}_k(\mbf{z}_1)^H, \cdots, \mbf{h}_k(\mbf{z}_B)^H ]^H \in \mbb{C}^{BN}$ denote the channel  from the $k$-th user to the 6DMA-BS by considering all its $B$ 6DMA surfaces. Following (\ref{hkb}), we have
\begin{align}
	\mbf{h}_k(\mbf{z}) &= \sum_{l=1}^{L_k} \mbf{{G}}(\mbf{z},\mbf{f}_{k,l}) \mbf{a}(\mbf{z},\mbf{f}_{k,l}) v_{k,l} 
	=  \tilde{\mbf{A}}(\mbf{z},\mbf{F}_k) \mbf{v}_k,\label{h2}
\end{align}
where $\mbf{G}(\mbf{z},\mbf{f}_{k,l}) \in \mbb{R}^{BN\times BN}$  is a diagonal matrix with its $(i,i)$-th entry representing the antenna gain of the $i$-th antenna, $i=1,\cdots,BN$, of all the $B$ 6DMA surfaces, given by 
\begin{align}
	\mbf{{G}}(\mbf{z},\mbf{f}_{k,l}) = \mathrm{diag}\left(\sqrt{g(\mbf{u}_1,\mbf{f}_{k,l})},\cdots,\sqrt{g(\mbf{u}_B,\mbf{f}_{k,l})}\right) \otimes \mbf{I}_N.
\end{align}
The matrix $\tilde{\mbf{A}}(\mbf{z},\mbf{F}_k) \in \mbb{C}^{BN\times L_k}$ resembles the  steering vectors of all paths of the $k$-th user weighted by antenna gains, given by
\begin{align}
	\tilde{\mbf{A}}(\mbf{z},\mbf{F}_k)  = \mbf{{G}}(\mbf{z},\mbf{f}_{k,l})\left[{\mbf{a}}(\mbf{z},\mbf{f}_{k,1}),\cdots,{\mbf{a}}(\mbf{z},\mbf{f}_{k,L_k})\right], \label{weighted_steering_matrix}
\end{align}
where $\mbf{a}(\mbf{z},\mbf{f}_{k,l}) \triangleq [\mathbf{{a}}(\mbf{z}_1,\mbf{f}_{k,l})^H, \cdots, \mathbf{{a}}(\mbf{z}_B,\mbf{f}_{k,l})^H]^H\in \mbb{C}^{BN}$,
$\mbf{F}_k  \triangleq [\mbf{f}_{k,1},\cdots,\mbf{f}_{k,L_k}]\in \mbb{R}^{3\times L_k}$, and  $\mbf{v}_k  \triangleq [v_{k,1},\cdots,v_{k,L_k}]^\top \in \mbb{C}^{L_k}$.

Assuming rich local scattering around each user, we model each channel vector $\mathbf v_k$ as independent circularly symmetric complex Gaussian (CSCG):
\begin{align}
	\mbf{v}_k \sim \mc{CN}(0,\mbf{D}_k),
\end{align}
where $\mbf{D}_k \triangleq \text{diag}(\alpha_{k,1}^2,\cdots,\alpha_{k,L_k}^2) \in \mbb{R}^{L_k\times L_k}$ with its $(l,l)$-th entry denoting the average power  of the $l$-th path of the $k$-th user to the BS.
In other words, the channels of users are independently sampled from 
\begin{align}
	\mbf{h}_k(\mbf{z})  \sim \mc{CN}(0,\mbf{\Sigma}_k(\mbf{z})), \label{hk_distri}
\end{align}
where the channel covariance matrix $\mbf{\Sigma}_k(\mbf{z}) \in \mbb{C}^{BN\times BN}$ is given by
\begin{align}
	\mbf{\Sigma}_k(\mbf{z}) = \tilde{\mbf{A}}(\mbf{z},\mbf{F}_k) \mbf{D}_k \tilde{\mbf{A}}(\mbf{z},\mbf{F}_k)^H. \label{R_def}
\end{align}
As a consequence of (\ref{hk_distri}) and (\ref{R_def}),  the statistical characteristics of all channels are fully determined by the position-rotation state of 6DMA-BS $\mbf{z}$. For this reason, we define $\mbf{\Sigma}(\mbf{z}) = [\mbf{\Sigma}_1(\mbf{z}),\cdots,\mbf{\Sigma}_K(\mbf{z})] \in \mbb{C}^{BN\times BNK}$  as the users' SCI, which is assumed to be known a priori.

\subsection{Unified Constraint for Blockage and Overlap Avoidance}
In this subsection, we introduce the unified constraint for 6DMA surfaces' placement, which prevents mutual signal reflections and physical overlap between any two 6DMA surfaces. First, we define the non-positive halfspace associated with the $b$-th 6DMA surface as
\begin{align}
	\mc{H}^-_b(\mbf{q}_b,\mbf{u}_b) = \{\mbf{x}\in\mbb{R}^3\mid \mathbf{n}(\mbf{u}_b)^{\top}(\mathbf{x}-\mathbf{q}_{b})\leq  0\},
\end{align}
where the  normal vector of the $b$-th 6DMA surface $\mbf{n}(\mbf{u}_b)$ is given by
\begin{align}
	\mbf{n}(\mbf{u}_b) = \mbf{R}(\mbf{u}_b)\bar{\mbf{n}} \label{n_def},
\end{align} 
where  $\bar{\mbf{n}}$ is the predefined normal vector of a 6DMA surface in its local CCS. Next, we define the  $b$-th 6DMA surface's occupied  region in the global CCS as 
\begin{align}
	\mbf{\mc{V}}_b(\mbf{q}_b,\mbf{u}_b) = \{\mbf{q}_b + \mbf{R}(\mbf{u}_b) \mbf{x}\mid \mbf{x}\in\mc{\bar{V}}\},
\end{align}
where $\mc{\bar{V}}$ denotes the predefined 6DMA surface  region in its local CCS. Finally, we confine the placement of each 6DMA surface  within the non-positive halfspaces associated with all the other 6DMA surfaces, with the corresponding constraint  given by
\begin{align}
	\mbf{\mc{V}}_{{b'}}(\mbf{q}_{{b'}},\mbf{u}_{{b'}}) \subseteq \mc{H}^-_{b}(\mbf{q}_{b},\mbf{u}_{b}),~\forall {b} ,{{b'}} \in \mathcal{B}, b\neq b'. \label{P_feasible_c11}
\end{align}


\subsection{Problem Formulation}
In this paper, we optimize the 6DMA position-rotation state vector $\mbf{z}$, for maximizing the average sum log-rate of the multi-user (uplink) multiple-access channel by taking into account the rate fairness among the users. Assuming that the minimum mean-square error (MMSE) based linear receiver is applied at the 6DMA-BS to detect the users' signals independently, the achievable average  rate of the $k$-th user can be written as
\begin{align}
	r_k(\mbf{\Sigma}(\mbf{z})) = \expa{\log_2 (1 + \mbf{h}_k^H \mbf{B}_k^{-1}\mbf{h}_k )}, \label{r_origin}
\end{align}
where the interference-plus-noise covariance matrix $\mbf{B}_k \in \mbb{C}^{BN \times BN}$ is given by
\begin{align}
	\mbf{B}_k = \sum_{k' \ne k} \frac{p_{k'}}{p_k}\mbf{h}_{k'}\mbf{h}_{k'}^H + \frac{\sigma^2}{p_k} \mbf{I}_{BN},\label{B_def}
\end{align}
where $p_k$ represents the transmit power value of the user $k$ and $\sigma^2$ denotes the average noise power at the BS's receiver. In (\ref{r_origin}), the expectation is taken over the random channels $\mathbf h_k$ of all users, conditioned on their SCI $\mathbf\Sigma(\mathbf z)$. To derive a more tractable expression, we apply Jensen’s inequality to upper-bound the average rate by
\begin{align}
	\bar{r}_k(\mbf{\Sigma}(\mbf{z}))  = \log_2  \left( 1 + \trace{\expa{\mbf{B}_k^{-1}   } \mbf{\Sigma}_k(\mbf{z})   }  \right) \label{r_ub}.
\end{align}
Note that computing $\expa{\mbf{B}_k^{-1}   }$ in (\ref{r_ub}) is still a non-trivial task, thus we again apply Jensen's inequality to lower bound (\ref{r_ub}) by
\begin{align}
	\underline{r}_k(\mbf{\Sigma}(\mbf{z})) = \log_2 \left( 1 + \trace{\expa{\mbf{B}_k }^{-1}   \mbf{\Sigma}_k(\mbf{z})  }  \right) , \label{r_ub_lb}
\end{align}
where  $\expa{\mbf{B}_k}$ can be easily derived as
\begin{align}
	\expa{\mbf{B}_k} =  \sum_{k' \ne k} \frac{p_{k'}}{p_k}\mbf{\Sigma}_{k'}(\mbf{z})  + \frac{\sigma^2}{p_k} \mbf{I}_{BN}.
\end{align}
By replacing the  average rate $r_k$ in (\ref{r_origin}) by the approximation $\underline{r}_k$, the optimization problem is finally  formulated as
\begin{subequations}
	\label{Sta_Opt}
	\begin{align}
	\!\!	\text{(P1)}~~&~\mathop{\max}\limits_{\mathbf{\mbf{z}}}~~
		\sum_{k=1}^{K} \log(\underline{r}_k(\mbf{\Sigma}(\mbf{z}))) \label{P1_obj}\\
		~&~	\mbf{\mc{V}}_{{b'}}(\mbf{q}_{{b'}},\mbf{u}_{{b'}}) \subseteq \mc{H}^-_{b}(\mbf{q}_{b},\mbf{u}_{b}),~\forall {b} ,{{b'}} \in \mathcal{B}, b\neq b', \label{P1c1}\\
		~&~ \mathbf{q}_{b} \in \mathcal{V}_{\text{6DMA}}, ~\forall b  \in \mathcal{B} \label{P1c3}.
	\end{align}
\end{subequations}
The main challenge in solving (P1) lies in the tight coupling between the position and rotation variables of the 6DMA surfaces in the objective function, which makes the problem difficult to solve optimally. 

\section{6DMA Position and Rotation Optimization Based on SCI}
In this section, we design a sequential optimization method to decouple position and rotation. Specifically, we optimize only the surface rotations $\mathbf u_b$ by expressing the positions $\mathbf q_b$ as functions of those rotations. Given the optimized rotations of 6DMA surfaces, we then determine their feasible positions subject to the placement constraints.
\subsection{Rotation Optimization}
In this subsection, we relax (P1) to an optimization problem over  the rotations of the 6DMA surfaces only.  First, we drop the unified blockage and overlap constraint (\ref{P1c1}). Note that the remaining constraint ensures that the feasible surfaces lie within the 6DMA region, while a spherical region provides a simple construction that satisfies this condition. To  confine the 6DMA surfaces to a spherical region, we set the position-rotation pair of each 6DMA surface as
\begin{align}
	\mbf{z}_b(\mbf{u}_b) = [d_{\text{ins}} \mbf{n}(\mbf{u}_b)^\top, \mbf{u}_b^\top]^\top, ~\forall b \in \mc{B}, \label{zb_over_ub}
\end{align}
where $d_{\text{ins}}$ denotes the largest possible radius of an inscribed sphere inside the 6DMA region, $\mc{V}_{\text{6DMA}}$. Accordingly, the  position-rotation state of the $B$ 6DMA surfaces takes the form
\begin{align}
	\mbf{z}(\mbf{u}) =[\mbf{z}_1(\mbf{u}_1),\cdots,\mbf{z}_B(\mbf{u}_B)], \label{z_over_u}
\end{align}
where $\mbf{u} \triangleq [\mbf{u}_1^\top,\cdots,\mbf{u}_B^\top]^\top\in\mbb{R}^{3B}$.
Substituting (\ref{z_over_u}) into (\ref{P1_obj}), (P1) is relaxed  into the following unconstrained optimization problem:
\begin{subequations}
	\label{P3A}
	\begin{align}
		\text{(P2-A)}~~&~\mathop{\max}\limits_{\mathbf{u}}~~
		\sum_{k=1}^{K} \log(\underline{r}_k(\mbf{\Sigma}( \mbf{z}(\mbf{u})))).
	\end{align}
\end{subequations}

Next, we apply a gradient descent algorithm to solve (P2-A). Denote the objective function in (P2-A) by $\tilde{r}(\mbf{u})$. The gradient of $\tilde{r}(\mbf{u})$ can be calculated numerically based on its definition. Specifically, define $\mbf{e}_i \in \mbb{R}^{3B}$ as a vector whose $i$-th entry is 1 and otherwise 0, and $\epsilon$ as a small positive number. Then, the partial derivative of $\tilde{r}(\mathbf{u})$ w.r.t. the $i$-th entry of $\mbf{u}$, denoted by $u_i$, $i=1,\cdots,3B$, is approximated by
\begin{align}
	\frac{\partial \tilde{r}(\mbf{u})}{\partial u_i} \approx \left(\tilde{r}(\mbf{u} + \epsilon   \mbf{e}_i) -\tilde{r}(\mbf{u}) \right)/\epsilon, \label{Nabla_1}
\end{align}
and the corresponding gradient is given by
\begin{align}
	\frac{\partial \tilde{r}(\mbf{u})}{\partial \mbf{u}} \triangleq [\frac{\partial \tilde{r}(\mbf{u})}{\partial {u}_1},\cdots,\frac{\partial \tilde{r}(\mbf{u})}{\partial {u}_{3B}}]^\top \label{Nabla_2}.
\end{align}
In each iteration, the descent direction of the rotations of the 6DMA surfaces is set to $\frac{\partial \tilde{r}(\mbf{u})}{\partial \mbf{u}}$, and the corresponding step size is determined by the Armijo rule \cite{armijo}. The proposed algorithm starts with an initial vector  $\mbf{u}^{(0)}$ (to be specified later) and generates a sequence of vectors $\{\mbf{u}^{(\kappa)}\}$, $\kappa=1,\cdots,\kappa_{\text{max}}$, as
\begin{align}
	\mbf{u}^{(\kappa)} = \mbf{u}^{(\kappa-1)} + \tau^{(\kappa-1)}\left. \frac{\partial \tilde{r}(\mbf{u})}{\partial \mbf{u}}\right|_{\mbf{u}^{(\kappa-1)}}.
\end{align}
The step size $\tau^{(\kappa-1)}$ in the $(\kappa-1)$-th iteration is calculated as $\tau_{\text{ini}} \delta^\nu$, where $\tau_{\text{ini}}$ denotes the initial step size, $\delta \in (0,1)$, and $\nu$ denotes the smallest nonnegative integer which satisfies
\begin{align}
	\tilde{r}(\mbf{u}^{(\kappa)}) - \tilde{r}(\mbf{u}^{(\kappa-1)}) >  \tau_{\text{ini}} \delta^\nu  \left. \frac{\partial \tilde{r}(\mbf{u})}{\partial \mbf{u}}\right|_{\mbf{u}^{(\kappa-1)}}^\top (\mbf{u}^{(\kappa)} - \mbf{u}^{(\kappa-1)}). \label{Armijo}
\end{align}

For initialization, we propose a greedy search algorithm to select the initial rotation vector, $\mbf{u}^{(0)}$, of the 6DMA surfaces to ensure good performance of the converged solution. First, let $\mc{C} \subset \mbb{R}^3$ denote a candidate set consisting of $\bar{M}$ rotations uniformly generated using the Fibonacci method. The initial rotation vector is then selected from $\mc{C}$ in an iterative manner. Specifically, at iteration $\kappa'$ (for $\kappa'=1,\dots,B$), let
$\hat{\mathbf u}^{(\kappa'-1)}\in\mathbb R^{3(\kappa'-1)}
$ denote the concatenation of the $\kappa'-1$ rotations chosen so far. In the $\kappa'$-th iteration, we perform an exhaustive search to identify the optimal $\hat{\mbf{u}}^* \in \mbb{R}^3$ from the candidate set $\mc{C}$ that maximizes the objective function of (P2-A), i.e.,
\begin{align}
	\hat{\mbf{u}}^* \leftarrow \arg\max_{\underline{\mbf{u}} \in \mc{C}} \tilde{r}([(\hat{\mbf{u}}^{(\kappa'-1)})^\top, \underline{\mbf{u}}^\top]^\top). \label{greedy_search_adaptation}
\end{align}
This procedure is repeated until all the rotations are selected for the $B$ 6DMA surfaces.

	\subsection{Position Optimization}
	Next, we aim to find feasible  positions of the 6DMA surfaces subject to their optimized rotations obtained in Section III-A. Denote the optimized rotations of 6DMA surfaces after solving (P2-A) by $\mbf{u}^* = [\mbf{u}_1^{*\top},\cdots,\mbf{u}_B^{*\top}]^\top \in \mbb{R}^{3B}$. In order to make the optimized rotations $\mbf{u}^*$ practically implementable, we formulate the following feasibility problem, i.e., 
	\begin{subequations}
		\label{feasible}
		\begin{align}
		\!\!\!\!	\text{(P2-B)}~&~\mathop{\text{Find}}~~\mbf{q}\in\mbb{R}^{3B}~~\text{such that}\nonumber\\
			&~\mc{V}_{{b'}}(\mbf{q}_{{b'}},\mbf{u}_{{b'}}^*) \subseteq \mc{H}_{b}^-(\mbf{q}_{b},\mbf{u}_{b}^*) ,~\forall {b} ,{{b'}} \in \mathcal{B}, b\neq b', \label{P_feasible_c1}\\
			&~ \mathbf{q}_{b} \in \mathcal{V}_{\text{6DMA}}, ~\forall b  \in \mathcal{B}, \label{P_feasible_c3}
		\end{align}
	\end{subequations}
	where $\mbf{q} \triangleq [\mbf{q}_1^\top,\cdots,\mbf{q}_B^\top]^\top\in\mbb{R}^{3B}$. 
	To cope with this non-convex (due to (\ref{P_feasible_c1})) optimization problem, we propose a geometry-based algorithm to search for the feasible positions of the $B$ 6DMA surfaces. 
	First, to simplify the positioning algorithm, we tighten constraint (\ref{P_feasible_c1}) by replacing it with
	\begin{align}
		\mc{V}^{\text{e}}_{{b'}}(\mbf{q}_{{b'}}) \subseteq \mc{H}_{b}^-(\mbf{q}_{b},\mbf{u}_{b}) ,~\forall {b} ,{{b'}} \in \mathcal{B}, b\neq b', \label{P_feasible_c1_new}
	\end{align}
	where the circular extended region (CER) of the $b$-th 6DMA surface, $\mc{V}^{\text{e}}_b(\mbf{q}_b)\subset\mbb{R}^3$, is defined as a circle centered at $\mbf{q}_b$, with normal vector $\mbf{n}^*_b$ and radius $d/2$, i.e.,
	\begin{align}
		\mc{V}^{\text{e}}_b(\mbf{q}_b) \triangleq \{\mbf{x}\in\mbb{R}^3\mid \|(\mbf{I}_3 - \mbf{n}^*_b \mbf{n}^{*\top}_b)(\mbf{x}-\mbf{q}_b)\|_2 \le \frac{d}{2}\} \label{CER_def}.
	\end{align}
	Note that the radius of the CER,  \( \frac{d}{2} \), is chosen as the minimum radius required to enclose the 6DMA surface region.  Since  $\mbf{\mc{V}}_b(\mbf{q}_b,\mbf{u}^*_b) \subseteq \mc{V}^{\text{e}}_b(\mbf{q}_b)$,    constraint (\ref{P_feasible_c1_new}) guarantees that a larger region of each 6DMA surface lies within the non-positive halfspace associated with all other surfaces.

	Next, we define $\mc{X}^{(\kappa)}$ and $\mc{Y}^{(\kappa)}$ as the index sets which include the indices of the already positioned  6DMA surfaces and the remaining non-positioned  6DMA surfaces at the $\kappa$-th iteration, $\kappa=1,\cdots,B$, respectively. Thus, $\mc{X}^{(0)} $ is an empty set and $\mc{Y}^{(0)} = \mc{B}=\{1,\cdots,B\}$.  
	At the first iteration, we randomly select a surface and assign its position arbitrarily, for example, the coordinate origin. At the $\kappa$-th iteration, we position the  surface whose normal vector is most closely aligned with those  in $\mc{X}^{(\kappa-1)}$. Specifically, the index of the surface to be positioned is given by
	\begin{align}
		b^{(\kappa)} = \arg\mathop{\max}_{b\in \mc{Y}^{(\kappa-1)}}  \{ \max_{{b'}\in \mc{X}^{(\kappa-1)}} {\mbf{n}^*_{b}}^\top \mbf{n}^*_{{b'}} \}.
	\end{align}
	Then, we jointly position the $b^{(\kappa)}$-th 6DMA surface along with the surfaces indexed by $\mc{X}^{(\kappa-1)}$, while ensuring that  the constraint ($\ref{P_feasible_c1_new})$ is satisfied. The positioning policy for the $b^{(\kappa)}$-th 6DMA surface consists of three steps in each iteration. Specifically, the operations of the $\kappa$-th iteration include: 
	\begin{itemize}[]
		\item{Step 1:} Select the hyperplane (denoted by $\mc{H}^{(\kappa)}$) on which the $b^{(\kappa)}$-th surface is to be placed. Specifically, the normal vector of $\mc{H}^{(\kappa)}$ coincides with that of the  $b^{(\kappa)}$-th 6DMA surface. Moreover, the CERs associated with 6DMA surfaces  in $\mc{X}^{(\kappa-1)}$ should lie within the non-positive halfspace of  $\mc{H}^{(\kappa)}$,  and  the  distance between these CERs and $\mc{H}^{(\kappa)}$ should be minimized. Thus, the hyperplane $\mc{H}^{(\kappa)}$ is given by
		\begin{align}
			\mc{H}^{(\kappa)} = \left\{\mbf{x}\in\mbb{R}^3\mid {\mbf{n}^{*\top }_{b^{(\kappa)}} } \mbf{x} = \chi^{(\kappa)} \right\}, \label{get_plane1}
		\end{align}
		where $\chi^{(\kappa)}$ is given by
		\begin{align}
			\chi^{(\kappa)}  &= \max_{b \in \mc{X}^{(\kappa-1)}} \max_{\mbf{x}\in \mc{V}^{\text{e}}_{b}(\mbf{q}_{b})}  {\mbf{n}^{*\top}_{b^{(\kappa)}} } \mbf{x} . \label{Compute_r}
		\end{align} 
		Equation (\ref{Compute_r}) ensures that at least one of the CERs associated with 6DMA surfaces  in $\mc{X}^{(\kappa-1)}$ is tangent to the hyperplane $\mc{H}^{(\kappa)}$. We denote the index of the  tangent CER and the position of its corresponding tangent point as
		\begin{align}
			b^{(\kappa)}_{\text{tan}},\mbf{x}_{\text{tan}}^{(\kappa)}&= \arg\mathop{\max}_{b \in \mc{X}^{(\kappa-1)}, \mbf{x}\in \mc{V}^{\text{e}}_{b}(\mbf{q}_{b})}  {\mbf{n}^{*\top}_{b^{(\kappa)}} } \mbf{x}.
		\end{align}

		\item{Step 2:} Select the position of the $b^{(\kappa)}$-th 6DMA surface. Since the $b^{(\kappa)}$-th surface is placed on $\mc{H}^{(\kappa)}$, the CERs associated with 6DMA surfaces  in $\mc{X}^{(\kappa-1)}$ are already  within its non-positive halfspace. In this step, we  select $\mbf{q}_{b^{(\kappa)}}$ such that $	\mc{V}^{\text{e}}_{b^{(\kappa)}}(\mbf{q}_{b^{(\kappa)}})$ is also within the non-positive halfspaces of these CERs. Denote the projection of the  normal vector of the  $b^{(\kappa)}_{\text{tan}}$-th 6DMA surface (which is tangent to the hyperplane $\mc{H}^{(\kappa)}$) onto  $\mc{H}^{(\kappa)}$, as
		\begin{align}
			\mbf{n}_{b^{(\kappa)}_{\text{tan}},\mc{H}}^*\triangleq \mbf{n}_{b^{(\kappa)}_{\text{tan}}}^* - \left({\mbf{n}^{*^\top}_{b^{(\kappa)}_{\text{tan}}}} \mbf{n}^*_{b^{(\kappa)}} \right)\mbf{n}^*_{b^{(\kappa)}}.
		\end{align} 
		Then, we obtain an initial guess for the position of the  $b^{(\kappa)}$-th 6DMA surface given by
		\begin{align}
			\mbf{q}_{b^{(\kappa)}}^{\text{ini}} = \mbf{x}_{\text{tan}}^{(\kappa)} - \frac{d}{2}\frac{\mbf{n}_{b^{(\kappa)}_{\text{tan}},\mc{H}}^*}{\|\mbf{n}_{b^{(\kappa)}_{\text{tan}},\mc{H}}^*\|_2}. \label{q_ini_def}
		\end{align}
		If $\mc{V}^{\text{e}}_{b^{(\kappa)}}(\mbf{q}_{b^{(\kappa)}}^{\text{ini}})$ lies within the non-positive halfspaces of the CERs associated with 6DMA surfaces  in $\mc{X}^{(\kappa-1)}$, we assign
		\begin{align}
			\mbf{q}_{b^{(\kappa)}}  &\leftarrow	\mbf{q}_{b^{(\kappa)}}^{\text{ini}}.
		\end{align}
		Otherwise, we shift each already positioned surface indexed by $\mc{X}^{(\kappa-1)}$ outward along its associated normal vector projected onto $\mc{H}^{(\kappa)}$ by a distance $d/2$ in order to make more space for positioning the $b^{(\kappa)}$-th 6DMA surface, i.e., 
		\begin{align}
			\mbf{q}_b  \leftarrow	\mbf{q}_b + \frac{d}{2}\frac{\mbf{n}^*_{b,\mc{H}}}{\|\mbf{n}^*_{b,\mc{H}}\|_2},~\forall b\in \mc{X}^{(\kappa-1)}, \label{q_update}
		\end{align}
		where the projected normal vector is given by
		\begin{align}
			\mbf{n}^*_{b,\mc{H}} \triangleq \mbf{n}_b^* - \left(\mbf{n}^{*\top}_b \mbf{n}^*_{b^{(\kappa)}} \right)\mbf{n}^*_{b^{(\kappa)}}.
		\end{align}
		Then, we assign
		\begin{align}
			\mbf{q}_{b^{(\kappa)}}  &\leftarrow	\mbf{x}_{\text{tan}}^{(\kappa)}. \label{q_update_2}
		\end{align}
		
		\newtheorem{lemma}{Lemma}
				
		\item{Step 3:} If $\kappa< B$, we update 
		\begin{subequations}
		\begin{align}
			\mc{X}^{(\kappa)} &\leftarrow  \mc{X}^{(\kappa-1)} \cup \{b^{(\kappa)}\} ,\\
			\mc{Y}^{(\kappa)} &\leftarrow  \mc{Y}^{(\kappa-1)} \setminus \{b^{(\kappa)}\},\\
			\kappa &\leftarrow \kappa+1,
		\end{align}
		\end{subequations}
		where $\setminus$ denotes the set difference operator.
		We then proceed with the next iteration; otherwise, we stop the algorithm and return the final positions of the $B$ 6DMA surfaces.
	\end{itemize}

	\section{Simulation Results}
In the simulation, we consider five users and three dominant scatterers. The 6DMA-BS is located at \((0, 0, 0)\), and the positions of the dominant scatterers are \((-40, 30, 10)\), \((20, 0, 10)\), and \((0, -10, 0)\), respectively, with all coordinates given in meter (m). Two users are located around \((-40, 50, 0)\), one user is located at \((30, 80, 0)\), and two users are located around \((-10, -20, 0)\), respectively. The path loss model is given by \( v = \left(\frac{\lambda}{4\pi}\right)^2 d^{-\eta} \), where the path loss exponent is \( \eta = 3 \), and the carrier wavelength is \( \lambda = 0.125\,\mathrm{m} \). The direct links between all users and the BS are assumed to be blocked, and all users share the same set of dominant scatterers. 
	The 6DMA-BS is equipped with \( B = 8 \) 6DMA surfaces, with each  equipped with an UPA  of \( N = 4 \) antennas. The edge length of each 6DMA (square) surface is \( \sqrt{2} \lambda \), and the positions of its four antennas, \( \bar{\mathbf{r}}_n \)'s, are located at \( [0, \pm \frac{1}{4}\lambda, \pm \frac{1}{4}\lambda] \) in its local coordinate system. The edge length of the cubic region where the $B$ 6DMA surfaces can be flexibly positioned/rotated,  \( \mathcal{V}_{\text{6DMA}} \), is set to \( \sqrt{2}\,\text{m} \). The antenna gain pattern in (\ref{antenna_gain}) follows the 3GPP standard \cite{3gpp2017}. We set \( \bar{M} = 512 \),  \( \kappa_{\max} = 20 \), and \( \epsilon = 2^{-16} \).
		\begin{figure}[!t] 
		\centering
		\setlength{\abovecaptionskip}{-3pt}
		\setlength{\belowcaptionskip}{-15pt}
		\includegraphics[width=3.3in]{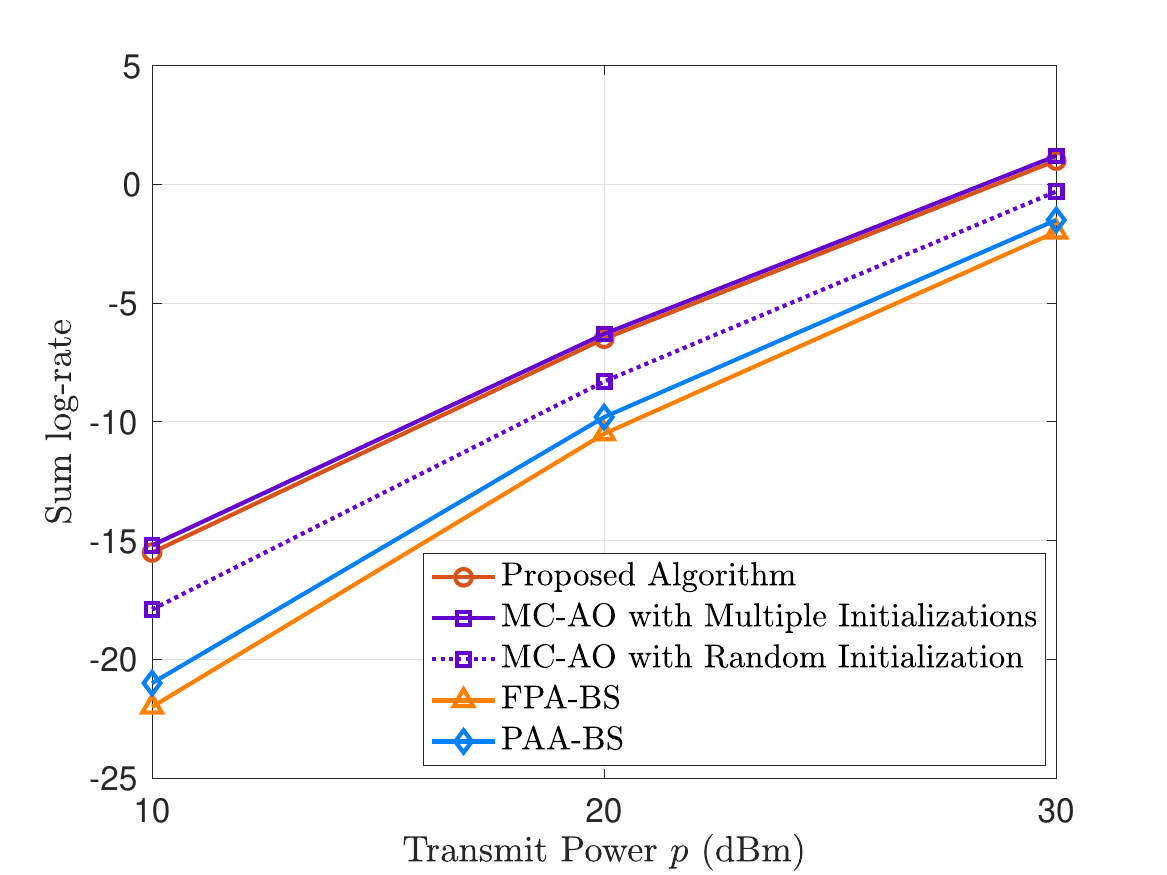} 
		\caption{Network  average sum log-rate versus transmit power  of users.} 
		\label{result_cvp} 
		\vspace{-0.58cm}
	\end{figure}
	
	For performance comparison, we consider the following benchmark schemes.
	\begin{enumerate}
			\item Monte Carlo with AO (MC-AO):   the expectation over $r_k$ is computed by the Monte Carlo method. 
		In this  scheme, the distribution of $\mbf{H}$ is assumed to be perfectly known, and the  number of channel realizations  is set to  $W=10^3$. In addition, the alternating optimization method proposed in \cite{6DMA_TWC} is used to optimize the 6DMA positions and rotations. 
		\item Fixed position BS (FPA-BS): a fixed three-sector BS is considered with each sector consisting of 11 fixed (position and rotation)  antennas (FPAs). The elevation tilt angle is set as 0. 
		\item Position-adjustable antenna aided BS (PAA-BS): a three-sector BS is considered, where each sector is equipped with 11 position-adjustable antennas (PAAs), which can move freely on each sector's two-dimensional (2D) surface with size $0.5\,\mathrm{m}\times0.5\,\mathrm{m}$ and inter-antenna minimum spacing equal to $\lambda/2$. The antenna positions are optimized using the particle swarm optimization (PSO) algorithm\cite{PSO} based on users' instantaneous channels.
	\end{enumerate}

In Fig. \ref{result_cvp}, we evaluate the network average sum log-rate versus user transmit power, assuming equal power for all users, i.e., $p_k = p$, $\forall k$. From Fig. \ref{result_cvp}, we observe that the MC-AO method with multiple initializations yields the best performance by accurately evaluating the expectation in (\ref{r_origin}), but suffers from high complexity and sensitivity to initialization. In contrast, our proposed algorithm achieves comparable performance with much lower complexity. The PAA-BS method slightly improves upon the FPA scheme by adjusting positions, but remains inferior to the proposed method.

\begin{figure}[!t] 
	\centering
	\setlength{\abovecaptionskip}{-3pt}
	\setlength{\belowcaptionskip}{-15pt}
	\includegraphics[width=3.3in]{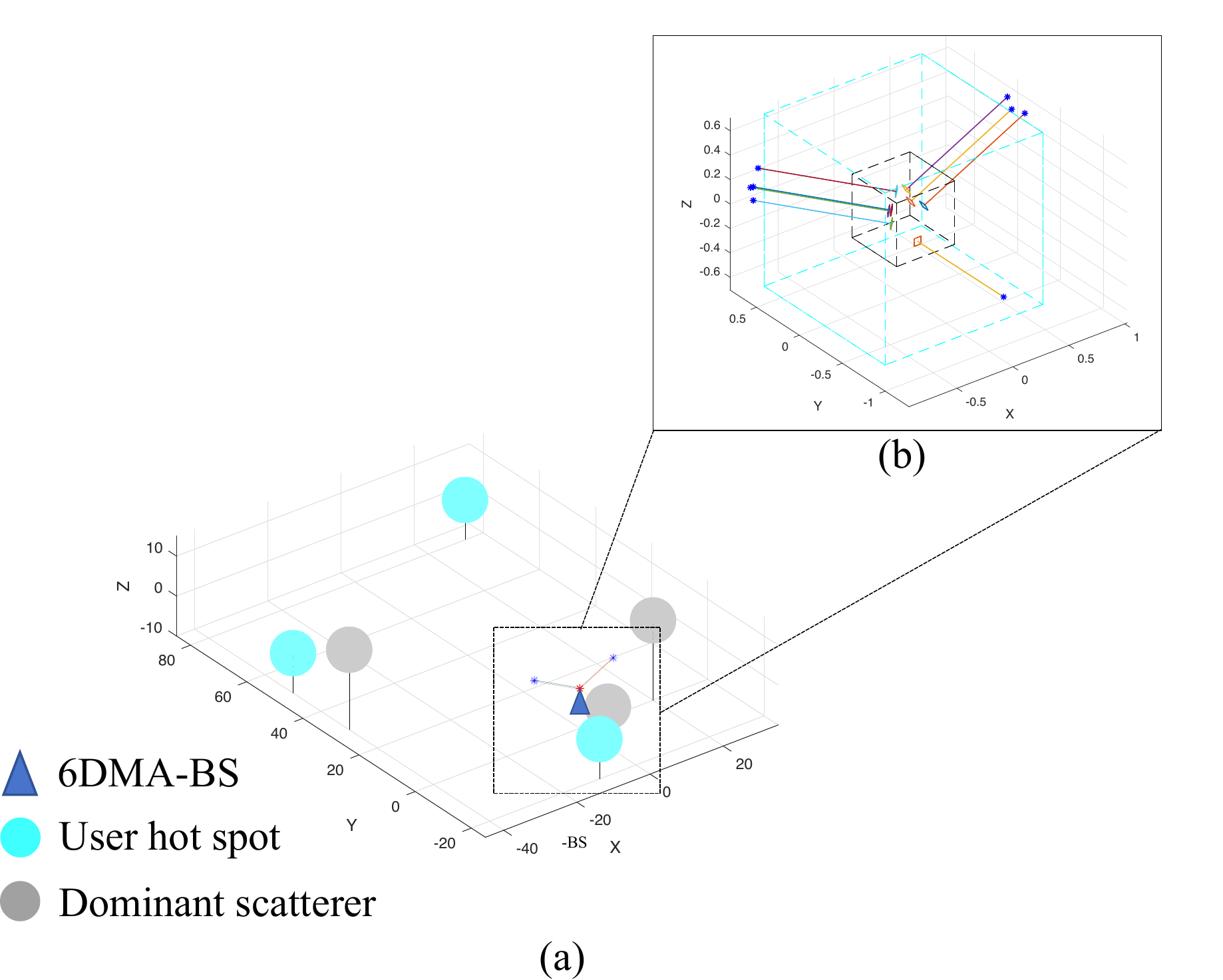} 
	\caption{Illustration of the positions and rotations  of 6DMA surfaces optimized by the proposed algorithm.} 
	\label{result_rp} 
	\vspace{-0.58cm}
\end{figure}

Fig. \ref{result_rp} illustrates the  optimized positions and rotations of 6DMA surfaces  by the proposed algorithm. Specifically, Fig.~\ref{result_rp}(a) visualizes the orientations of the 6DMA surfaces in the global coordinate system, with solid lines and star markers indicating their optimized normal vectors, while Fig.\,\ref{result_rp}(b) shows the optimized surface positions and rotations in the 6DMA-BS region. In Fig.\,\ref{result_rp}(b), all  6DMA surfaces are placed within a cube (indicated by black dash lines) with  edge length $d' = 0.52\,\mathrm{m}$, which is much smaller than the  edge length of $\mc{V}_{\text{6DMA}}$ ($\sqrt{2}\,\mathrm{m}$), indicating the effectiveness of the proposed method for finding the feasible 6DMA surfaces' positions given their designed rotations within the 6DMA-BS region.  

	\section{Conclusion}
This paper presented a practical low-complexity design framework for 6DMA-enabled communication systems. We formulated an optimization problem to maximize the average sum log-rate of a multi-user multiple-access channel based on  SCI. We  proposed a low-complexity sequential optimization method that first determines the optimal antenna surfaces' rotations and subsequently identifies their feasible positions to realize these rotations under practical antenna placement constraints. Simulation results demonstrated that our proposed algorithm based on SCI achieves performance close to the MC-AO benchmark, while significantly reducing its computational complexity.

	\bibliographystyle{IEEEtran}
	\bibliography{6DMA_bib}
\end{document}